\documentclass{optica-article}

\journal{opticajournal} % for journals or Optica Open

\articletype{Research Article}

\usepackage {mathrsfs} 
\usepackage{physunits}
\usepackage{siunitx}
\usepackage[numbers,sort&compress]{natbib}
\setcitestyle{numbers,square}
\usepackage{graphicx}
\usepackage{algorithm}
\usepackage{caption}
\usepackage{enumitem}
\usepackage{algpseudocode}
\usepackage{textcomp}

\usepackage{lineno}
\algnewcommand\algorithmicparfor{\textbf{for}}
\algnewcommand\algorithmicpardo{\textbf{do each in parallel independently}}
\algdef{S}[FOR]{ParFor}[1]{\algorithmicparfor\ #1\ \algorithmicpardo}
\begin{document}

\title{Subpixel correction of diffraction pattern shifts in ptychography via automatic differentiation}

\author{Zhengkang Xu,\authormark{1,2,$\dag$} Yanqi Chen,\authormark{2,$\dag$} Hao Xu,\authormark{1} Qingxin Wang,\authormark{2} Jin Niu,\authormark{1,2}  Lei Huang,\authormark{1,2} Jiyue Tang,\authormark{1,2} Yongjun Ma,\authormark{1} Yutong Wang,\authormark{1} Yishi Shi,\authormark{1,2,3} Changjun Ke,\authormark{1,2} Jie Li,\authormark{1,2,4} and Zhongwei Fan\authormark{2,*}}

\address{\authormark{1}Aerospace Information Research Institude, Chinese Academy of Sciences, Beijing, 100094, China\\
\authormark{2}School of Optoelectronics, University of the Chinese Academy of Sciences, Beijing, 100049, China\\
\authormark{3}Center for Materials Science and Optoelectronics Engineering, University of Chinese Academy of Sciences, Beijing, 100049, China\\
%\authormark{4}College of Mathematics and Physics, Hebei University of Engineering, Hebei, 056038,China\\
\authormark{$\dag$}The authors contributed equally to this work.\\
\authormark{4}e-mail: lijie430@aircas.ac.cn\\
\authormark{*}fanzhongwei@ucas.ac.cn}

% \email{\authormark{*}fanzhongwei@ucas.ac.cn} %% email address is required; see note below about the corresponding author designation

% use {asbstract*} to suppress the copyright line. Copyright information will be added in production

\begin{abstract*} 
Ptychography, a coherent diffraction imaging technique, has become an indispensable tool in materials characterization, biological imaging, and nanostructure analysis due to its capability for high-resolution, lensless reconstruction of complex-valued images. In typical workflows, raw diffraction patterns are commonly cropped to isolate the valid central region before reconstruction. However, if the crop is misaligned from the diffraction pattern's zero-order, reconstruction may suffer from slower convergence, phase wrapping, and reduced image fidelity. These issues are further exacerbated in experimental configurations involving reflective geometries or broadband illumination, where incorrect cropping introduces systematic preprocessing errors that compromise the entire ptychographic inversion. To address this challenge, we present an approach based on automatic differentiation (AD), where the cropping shift is treated as an optimizable parameter within the reconstruction framework. By integrating shift correction into the backpropagation loop, our method simultaneously refines the object, probe, and shift positions without requiring manual tuning. Simulation results demonstrate that, even with initial offsets ranging up to $\pm 5$ pixels, the proposed method achieves subpixel correction, with an average deviation below 0.5 pixels. Experiments in the extreme ultraviolet (EUV) regime further validate the method's robustness and effectiveness. This AD-based strategy enhances the automation and robustness of ptychographic reconstructions, and is adaptable to diverse experimental conditions.

\end{abstract*}

%%%%%%%%%%%%%%%%%%%%%%%%%%  body  %%%%%%%%%%%%%%%%%%%%%%%%%%
\section{Introduction}
Ptychography is an advanced coherent diffraction imaging (CDI) technique that allows the retrieval of both the complex illumination field and the function of an object through the measurement of overlapping diffraction patterns \cite{hawkes2019springer}. By scanning a localized, coherent probe across a sample, ptychography utilizes redundancy in the diffraction intensities from adjacent scanning positions to overcome the typical limitations of conventional CDI methods, such as ambiguities and inability to reconstruct the probes \cite{thibault2013reconstructing}. One of the major advantages of ptychography lies in its ability to achieve high-resolution imaging without relying on diffraction-limited optics or high-quality objective lenses. This makes it particularly attractive in wavelength regimes where traditional optical elements are difficult to fabricate or suffer from poor performance, such as in the extreme ultraviolet (EUV) range \cite{sandberg2007lensless, shao2024wavelength, eschen2024structured,eschen2023high, senhorst2024mitigating}. Owing to its inherent advantages—including high imaging resolution and stable reconstructions—and accompanied by methodological advances, ptychography has found increasing application across diverse scientific fields. These include X-ray and electron microscopy \cite{sandberg2007lensless, humphry2012ptychographic}, biomedical imaging \cite{shapiro2005biological, nelson2010high, giewekemeyer2010quantitative, kimura2014imaging}, and materials science \cite{PhysRevLett.110.205501, clark2013ultrafast, pfeifer2006three}.

To meet the demands of varying imaging regimes and experimental setups, ptychography has evolved into advanced formulations that more precisely align mathematical models with physical wave propagation. For instance, in the EUV regime, many optical systems rely on reflective rather than transmissive components. Since shallow incidences increase the reflected intensity, the incidence angles for EUV reflection measurements on semiconductor samples are often set close to grazing, generally between 70° and 80° with respect to the surface normal \cite{senhorst2024mitigating}. This grazing geometry introduces significant distortions in the recorded diffraction patterns, complicating the subsequent phase retrieval and image reconstruction processes. To address these distortions, the method called tilted-plane correction (TPC) has been widely adopted \cite{gardner2012high, lu2023characterisation, marathe2010coherent, porter2017general, seaberg2014tabletop}. TPC is a preprocessing approach that corrects the measured diffraction data prior to reconstruction by mapping the distorted curved grid onto a rectilinear grid. This correction enables the application of fast Fourier transform (FFT)-based algorithms for efficient phase retrieval and object reconstruction. Building upon the TPC framework, aPIE \cite{de2022apie} further enhances reconstruction quality by employing stochastic search strategies to mitigate resolution degradation caused by angular deviations.

The partial coherence of the illumination wavelength introduces significant challenges to accurate image reconstruction. Due to the inherent chromaticity of diffractive optics, the resulting diffraction patterns undergo wavelength-dependent spatial scaling \cite{chen2024ultra}. Consequently, the spectral broadening of the illumination leads to diffraction aliasing, which compromises the convergence of CDI algorithms. To address this issue, Fienup first introduced broadband CDI (BCDI) \cite{Fienup:99}, opening a new avenue for characterizing broadband radiation using multi-wavelength imaging, which is valuable with limited number of wavelength channels. Imaging with a partially coherent wavefront can be regarded as a blind deconvolution problem involving multiple discrete spectral components, where the mixed states of deconherence can be deconvolved by advanced reconstruction techniques, such as ptychographic information multiplexing (PIM) \cite{batey2014information, loetgering2021tailoring} and multi-wavelength techniques \cite{PhysRevA.79.023809, yao2021broadband}. An alternative approach, polyCDI, has also been proven to be an effective method with a pre-measured spectrum \cite{abbey2011lensless}.  More recently, numerical monochromatization algorithms have gained increasing attention for their capacity to tackle ultra-broadband spectral challenges \cite{huijts2020broadband, liu2023broadband, shearer2025robust, chen2024ultra}. This spectrum-dependent computational technique enables the extraction of quasi-monochromatic information from broadband coherent diffraction patterns through regularized matrix inversion. Owing to its role as a preprocessing operation, this method can be applied to any technique based on coherent elastic scattering of any type of coherent radiation \cite{huijts2020broadband}.

These complications, whether stemming from chromatic aberrations or geometric distortions, reflect the persistent discrepancies between computational models and physical measurements, as discussed above. As such, considerable research efforts have been devoted to minimizing these discrepancies. A key aspect of this endeavor involves the preprocessing of raw diffraction patterns acquired by detectors. This preprocessing typically includes background noise subtraction and the alignment of the central diffraction order (zero-order) with the center of the cropped diffraction frame. Accurate alignment is essential for reducing model-data inconsistencies and plays a pivotal role in suppressing reconstruction artifacts, phase wrapping errors, and numerical instabilities during iterative phase retrieval. The impact of misalignment becomes especially pronounced in complex imaging regimes such as reflection ptychography and numerically monochromatized broadband datasets. In these cases, even minor inaccuracies in zero-order alignment can result in substantial degradation of reconstruction quality, introducing not only translational shifts but also erroneous structural information, as elaborated in Section 2.2. Despite its importance, central alignment is often performed manually or via heuristic-based methods that rely on thresholding and researcher intuition. These conventional techniques are inherently limited in robustness, reproducibility, and scalability, underscoring the urgent need for more automated, stable, and precise solutions in ptychographic workflows.

In recent years, automatic differentiation (AD) has emerged as a transformative tool in computational imaging and has recently been integrated into various ptychographic frameworks \cite{ghosh2018adp, kandel2019using, seifert2021efficient, seifert2023maximum, shao2024wavelength, senhorst2024mitigating}. AD facilitates the exact and efficient computation of gradients with respect to all involved variables, obviating the need for hand-derived update rules when physical models are modified. This capability greatly accelerates the development and adaptation of ptychographic algorithms to evolving experimental conditions and imaging geometries.

Building upon this flexibility, we propose an AD-enabled optimization framework to directly address a critical factor affecting reconstruction fidelity: the lateral shift of recorded diffraction patterns. Unlike conventional preprocessing methods that treat these shifts as fixed or externally corrected, our approach incorporates them as learnable variables within the optimization loop and performs sub-pixel level correction during reconstruction. By jointly optimizing the object, probe, shift parameters, and other system-specific variables, our method yields a more accurate and physically consistent reconstruction. Importantly, this framework is designed to be broadly compatible with diverse ptychographic modalities—including transmission, reflection, and broadband (or multi-wavelength) configurations—without requiring model-specific derivations or case-dependent heuristics. The effectiveness of our approach is demonstrated through both numerical simulations and experimental validations, highlighting its potential for practical applications in high-precision ptychographic imaging.

\section{Theory}

\subsection{Forward model}

Far-field ptychography is an imaging technique that employs coherent light to scan a sample at multiple positions, capturing diffraction patterns at each scan point. This approach enables high-resolution reconstruction of the object's structure by solving an inverse problem that links the observed diffraction patterns to the underlying object properties.

In the typical setup of far-field ptychography, an unknown object is illuminated with a focused probe beam, producing diffraction patterns recorded at a detector plane. The object is modeled as $O \in \mathbb{C}^N$ and probe is modeled as $P \in \mathbb{C}^M$, where $N = N_{x} \times N_{y}$ and $M = M_{x} \times M_{y}$ are the total number of pixels of object and probe, respectively. A sequence of $K$ diffraction patterns are caputured, with the scanning positions described by a set of translation verctors $\boldsymbol{r} \in \mathbb{R}^{K \times 2}$. In the case of two parallel planes, with the paraxial and projection approximations  \cite{born2013principles}, the exit wave, $\Psi_{k}=P \cdot O({\mathbf{r}_k}) \in \mathbb{C}^M$, propagates to the detector and the corresponding intensity is given as:

\begin{figure}[h!]
	\centering
	\includegraphics[width=1\linewidth]{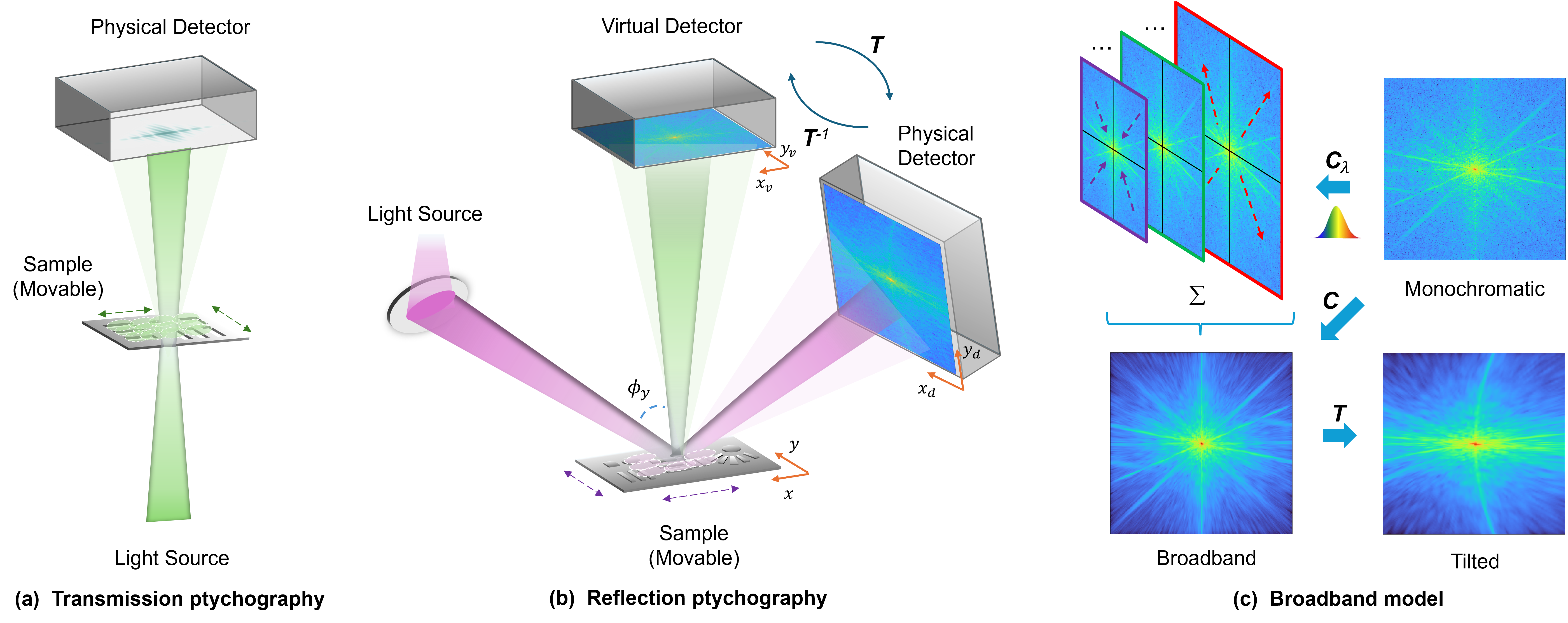}
	\caption[principle]{Schematic representations of the models under investigation. (a) Transmission ptychography. (b) Reflection ptychography. (c) Broadband model with optional incorporation of the reflection component. }
	\label{fig:fig1}
\end{figure}

\begin{equation}\label{eq1}
I_k=\left|\mathcal{F}\left\{\Psi_{k} \right\}\right|^2 + I_{b}
\end{equation}
where $\mathcal{F}$ is the propagation operator, $|\cdot|$ is the element-wise modulus of a vector and $I_{b}$ is the experimental background with positive components. The schematic diagram of the propagation process is shown in Fig.~\ref{fig:fig1} (a).

In reflection ptychography model, as shown in Fig.~\ref{fig:fig1} (b), the sample plane and the detector plane are typically not parallel, necessitating a modeling of light propagation between tilted planes. Accurately accounting for this angular misalignment is crucial for reconstructing the object from the experimentally acquired diffraction patterns on the detector. Conventionally, the diffraction pattern curvature is corrected during preprocessing stage by interpolating the collected diffraction patterns, which mitigates the curvature caused by tilted planes and simplifies the subsequent reconstruction. This procedure is known as tilted-plane correction (TPC) \cite{senhorst2024mitigating}.

Due to the nonlinearity of the transformation, the output grid exhibits irregular spacings, which is manifested on the diffraction pattern shown by the physical detector plane in in Fig.~\ref{fig:fig1} (b). However, fast Fourier transform (FFT) algorithms, which are widely used in ptychographic reconstruction, require data sampled on a uniform grid. Therefore, the measured intensities on the warped grid are interpolated onto a regular grid, forming what is referred to as the virtual detection plane in Fig.~\ref{fig:fig1} (b). The coordinate transformation between the physical detector coordinates $(x_d, y_d)$ and the virtual plane coordinates $(x_v, y_v)$ is described by the mapping:

\begin{equation}\label{eq2}
\boldsymbol{T}: x_v=\frac{x_d}{r_0} \cos \phi_y +{\sin \phi_y}\left[\left(1-\frac{x_d^2+y_d^2}{r_0^2}\right)^{1 / 2}-1\right], \quad y_v=\frac{y_d}{r_0}
\end{equation}
where $r_0=\sqrt{x_d^2+y_d^2+z^2}$ denotes the distance from the sample plane origin to a point in the detector plane. $z$ is the distance from the sample plane origin to the detector plane, and $\phi_y$ is the angle between the sample surface normal and the optical axis.

In the broadband model \cite{huijts2020broadband}, when polychromatic light is incident on a non-dispersive sample (i.e., the refractive indices of the materials in the sample do not change considerably over the spectrum of the source or these changes are spatially homogeneous at the reconstructed length scales), the diffraction pattern is the incoherent, spectrally weighted sum of the monochromatic diffraction patterns corresponding to all wavelengths present in the source. These monochromatic patterns are identical except for a geometric scaling.

As shown in Fig.~\ref{fig:fig1} (c), a monochromatic diffraction pattern at a reference central wavelength $\lambda^*$ can be transformed into diffraction patterns at other wavelengths through respective linear transformation $\boldsymbol{C}_{\lambda}$. Intuitively, this corresponds to a scaling effect: diffraction patterns at wavelengths shorter than $\lambda^*$ appear contracted, while those at longer wavelengths appear expanded. The broadband diffraction pattern is obtained by incoherent superposition of diffraction patterns of various wavelengths. This entire process can be depicted as the mapping process of $\boldsymbol{C}$. In the case of reflection ptychography model, an additional nonlinear transformation $\boldsymbol{T}$ must be considered. Therefore, the broadband model can ultimately be expressed in the following form:

\begin{equation}\label{eq3}
\mathbf{b}=\boldsymbol{T}(\boldsymbol{C} (\mathbf{m})) .
\end{equation}
where the broadband pattern is represented by the vector $\mathbf{b}$ and the monochromatic pattern is represented by the vector $\mathbf{m}$.

\subsection{Impact of offset}

In transmission ptychography model, a common preprocessing step involves cropping the diffraction pattern centered at the zero-order diffraction peak from the detector data to better align with the theoretical model described by Eq. (\ref{eq1}). However, any misalignment or shift in the cropped diffraction pattern introduces a model mismatch between the acquired physical data and the assumed computational model. This discrepancy can lead to slower convergence during the reconstruction process and artifacts such as phase wrapping in the recovered complex-valued object.

\begin{figure}[h!]
	\centering
	\includegraphics[width=1\linewidth]{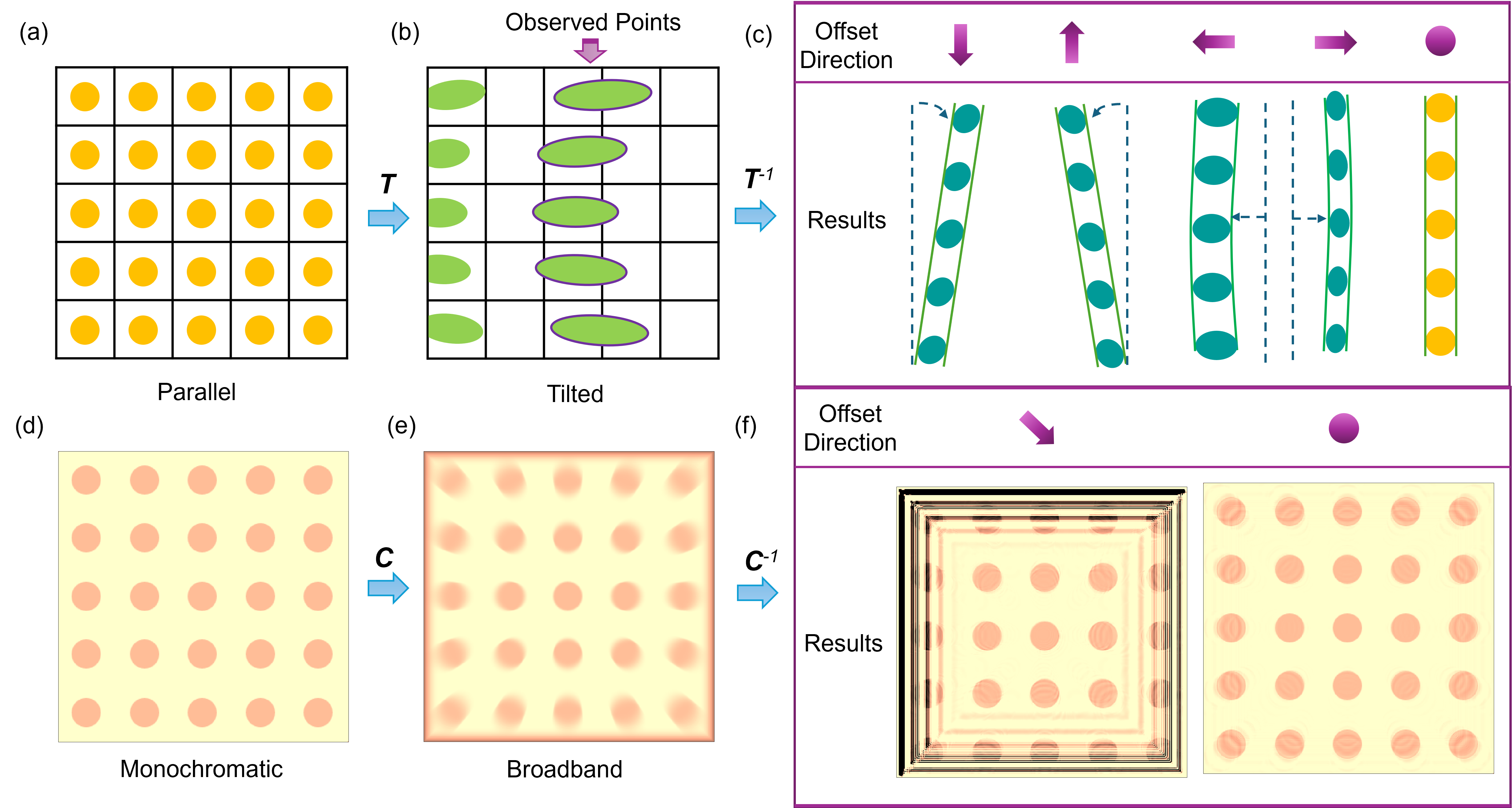}
	\caption[principle]{Illustration of the impact of diffraction pattern shifts on the preprocessing results of the reflection and broadband models. A diffraction pattern on a parallel detection plane is shown in (a), which is transformed into a tilted plane pattern (b) via $\boldsymbol{T}$ transformation. The result of the inverse transformation (TPC) is shown in (c), where the first row indicates the direction of the shift in the diffraction pattern center (arrows denote the direction of the shift; dot represents the unshifted reference), and the second row presents the corresponding inverse-transformed results under the given shifts. (d) shows a monochromatic diffraction pattern, which is converted into a broadband diffraction pattern (e) via $\boldsymbol{C}$ transformation, and then numerical monochromatization is applied to obtain the result in (f). Similarly, in (f), the first row indicates the direction of the diffraction pattern shift (with arrows and reference dots), and the second row shows the corresponding inverse-transformed results under various shift conditions.}
	\label{fig:fig2}
\end{figure}

The impact of diffraction pattern misalignment is even more pronounced in reflection ptychography model. As illustrated in  Fig.~\ref{fig:fig2} (a)-(c), the pattern of parallel virtual plane undergoes a non-linear transformation $\boldsymbol{T}$, projecting it onto the physical detector plane. When the input diffraction pattern is accurately centered, the inverse transformation $\boldsymbol{T}^{-1}$ can precisely recover vertically aligned circular features, as shown by the orange dots in Fig.~\ref{fig:fig2} (c), which correspond well with the ideal reference in Fig.~\ref{fig:fig2} (a). However, if the input pattern to the $\boldsymbol{T}^{-1}$  transform is offset, the non-linearity of the transformation distorts both the position and shape of the circular features, which is shown as green dots in  Fig.~\ref{fig:fig2} (c). Unlike transmission model, where shifts primarily cause displacement errors, in reflection ptychography, diffraction pattern misalignment leads to structural distortions in the preprocessed image. This significantly complicates the reconstruction process and poses a major challenge to algorithmic stability and accuracy.

A similar sensitivity to misalignment is observed in broadband ptychographic models, as demonstrated in Fig.~\ref{fig:fig2} (d)-(f). In this scenario, an ideal monochromatic diffraction pattern is transformed by a linear operator $\boldsymbol{C}$ into the broadband diffraction pattern recorded by the detector. To retrieve the constituent monochromatic components, Eq. (\ref{eq3}) must be solved (note that $\boldsymbol{T}$ is not involved here). These reconstructed monochromatic patterns serve as inputs for standard ptychographic algorithms, which typically assume quasi-monochromatic illumination. However, as with the $\boldsymbol{T}$ transformation, shifts in the recorded broadband pattern alter the input to the $\boldsymbol{C}$ mapping, thereby degrading the accuracy of the numerically retrieved monochromatic patterns. This distortion is evident in Fig.~\ref{fig:fig2} (f), where a lower-right shift in the broadband input leads to peripheral errors and central artifacts in the reconstructed image. In contrast, when no misalignment is present, the reconstruction more closely resembles the ideal case—though some edge ambiguity persists due to information loss during numerical monochromatization \cite{liu2023broadband}.

\subsection{Optimization}

Automatic Differentiation (AD) provides a powerful and general framework for solving optimization problems, including in the context of ptychography. By expressing the cost function using elementary mathematical operations with known derivatives, AD enables the automated computation of exact gradients via the chain rule. This eliminates the need for manual derivation of update rules, which is typically required in traditional PIE-based algorithms. It is a time-consuming task which has to be repeated after each modification of the forward model.

By recasting the reconstruction as an optimization problem over a set of global parameters, denoted as $\boldsymbol{\theta}$, we leverage AD to compute the exact gradients of the objective function without requiring closed-form derivations or approximations. Similar to the loss functions described in \cite{shao2024wavelength} and \cite{senhorst2024mitigating}, $\boldsymbol{\theta}$ includes not only the primary reconstruction targets—the probe $P$ and the object $O$—but also correction parameters that compensate for discrepancies between the forward model and the experimental conditions. The loss function for a single scanning position is based on the L2 loss and is defined as:

\begin{equation}\label{eq4}
\mathcal{L}\left(\boldsymbol{\theta}\right)=\sum_{k=1}^N \left(\sqrt{I_k}-\sqrt{\tilde{I}_k(\boldsymbol{\theta})} \right)^2 +\mathcal{R}
\end{equation}
where $I_k$ represents the measured intensity, and $\tilde{I}_k$ is the reconstructed intensity at the $k$-th scan position. $\mathcal{R}$ represents the regularization term with TV regularization. $\boldsymbol{\theta}=(P,O,\boldsymbol{\delta}, \boldsymbol{\xi}, \boldsymbol{\phi}, I_b)$ refers to model parameters, where $P$ and $O$ are probe and object, $\boldsymbol{\delta}$ represents the deviation of the probe scanning position, $\boldsymbol{\xi}$ represents the deviation of the diffraction pattern, $\boldsymbol{\phi}=(\phi_x,\phi_y,\phi_z)$ is the angles of the sample plane in the $x,y$ and $z$ coordinates relative to the actual orientation, and $I_b$ denotes the background noise and error that remain uncorrected in the image.

The entire algorithm is summarized in Alg. \ref{alg:tilt_optimization}. The fifth line of Alg. \ref{alg:tilt_optimization} corresponds to the main focus of the paper, which introduces optimization variables for controlling diffraction image offset correction. Note that, depending on the specific forward propagation model, the operators $\boldsymbol{T}$ and $\boldsymbol{C}$ can be optionally included. Additionally, the variable $\boldsymbol{\phi}$ depends on the presence of the $\boldsymbol{T}$ operator, whereas the variables $\boldsymbol{\xi}$ and $I_b$ are independent of both operators. The function $f$ represents the incorporation of optimization variables into the model. In the simulation studies, we focus on optimizing the probe $P$, the object $O$ alongside the newly iteratively optimized diffraction pattern shift $\boldsymbol{\xi}$. This is to specifically evaluate the performance of the diffraction shift optimization while minimizing potential interference from the optimization of other variables. In the experimental section, we demonstrate that our model is capable of jointly optimizing all model parameters simultaneously.

In this study, we adopt the Adam (Adaptive Moment Estimation) optimizer \cite{kingma2014adam}, an extension of the RMSProp algorithm that computes running averages of both the gradients and their second moments. Following standard terminology in the machine learning community, we define the minibatch size $b$ as the number of scan points used for a single gradient update. In the conventional ePIE algorithm, the object is updated after computing gradients from each individual scan point, corresponding to a minibatch size of $b=1$. In contrast, we set $b=16$ in our implementation, enabling parallel gradient computation across multiple scan points and applying a collective update to the object. For automatic differentiation and gradient computation, we use TensorFlow \cite{abadi2016tensorflow}, a Python-based library developed by Google, which is widely used for building and optimizing complex deep neural network architectures.

\begin{algorithm}[H] % [H] indicates that the algorithm is forcibly placed here.
\caption{An epoch of automatic differentiation-based diffraction pattern shift optimization}
\label{alg:tilt_optimization}
\begin{algorithmic}[1] % [1] indicate that the row numbers are displayed for each row. If row numbers are not required, they can be omitted.
    \Require{$I_d, \boldsymbol{r}$} \Comment{Measured intensities and scanning positions}
    \Require{$\boldsymbol{\theta} = \{P, O, \boldsymbol{\delta}, \boldsymbol{\xi}, \boldsymbol{\phi}, I_{b}\}$} \Comment{Initial parameters}
	\Require{$lr, N, B, b$} \Comment{Learning rates, \# iterations, \# batches, minibatch}

    \For{$n=1$ to $N$} 

	\State $\tilde{\boldsymbol{r}} \leftarrow f(\text{shuffle}(\boldsymbol{r}),\boldsymbol{\delta})$ \Comment{Randomize order}

        \ParFor{$b=1$ to $B$,}  \Comment{Computed in parallel for each $b$}

            \State $I_{v,k} \leftarrow \mathcal{F}(P \cdot O({\tilde{\boldsymbol{r}}_k}))$
            \State $\hat{I}_{d,k} \leftarrow f(\boldsymbol{T}(\boldsymbol{C} ({I_{v,k}})), \boldsymbol{\xi}, \boldsymbol{\phi}) + I_b$  \Comment{Add variables for pattern offset correction}
            \State $\partial_{\theta,k} \leftarrow \nabla_{\theta} \left[ \mathcal{L} \left( \sqrt{I_{d,k}}, \sqrt{\hat{I}_{d,k}} \right) + \mathcal{R} \right]$ \Comment{Computed using AD}
        \EndFor
        
        \State $\Delta \theta \leftarrow \text{ADAM}(\text{mean}(\partial_{\theta,k}))$ \Comment{Adjust update step by ADAM}
        \State $\theta \leftarrow \theta + \Delta \theta$
    \EndFor
\end{algorithmic}
\end{algorithm}

\section{ Results and discussion }

To validate the performance of the proposed algorithm and demonstrate its ability to generate reliable results by optimizing diffraction image shift errors, we conduct reconstructions using both simulated and experimental data.

\subsection{ Simulation }

In the simulation experiments, we evaluate the performance of the proposed algorithm using three distinct forward propagation models: the reflection model, the broadband model, and the reflection-broadband model. A resolution target (USAF 1951) is employed as the sample for reconstruction. The following simulation parameters are shared across all models: the propagation distance from the sample to the detector plane is set to $z=$ 25 mm; the size of the illuminating probe is $N=256 \times 256$ pixels; and the number of scanning positions is $K=$ 100. Model-specific parameters are configured as follows: for the reflection model, the incident angle is set to $\phi_y=$ 70°; for the broadband model, the spectral bandwidth is set to $\Delta \lambda / \lambda = 20 \%$. The reflection-broadband model incorporates both conditions, adopting the same incident angle and spectral bandwidth as the reflection and broadband components, respectively. To simulate subpixel-level displacements in the centers of the diffraction patterns, we introduce an offset $(x_{\text{offset}}^{i}, y_{\text{offset}}^{i})$ for each diffraction pattern, where $i=1,2,...,K$. These offsets are sampled from a uniform distribution within the range $(-a, a)$, where $a$ denotes the maximum pixel displacement.

Regarding the optimization parameters, we use the Adam optimizer during the optimization process. The learning rate is set to 0.1 for object reconstruction and 0.95 for pattern offset updates. To mitigate potential instability caused by large gradients during the initial iterations, the reconstruction is first conducted for 5 epochs without applying offset correction, which is activated after this initial phase. Furthermore, regularization constraints are not applied during the optimization in the simulation.

\begin{figure}[h!]
	\centering
	\includegraphics[width=1\linewidth]{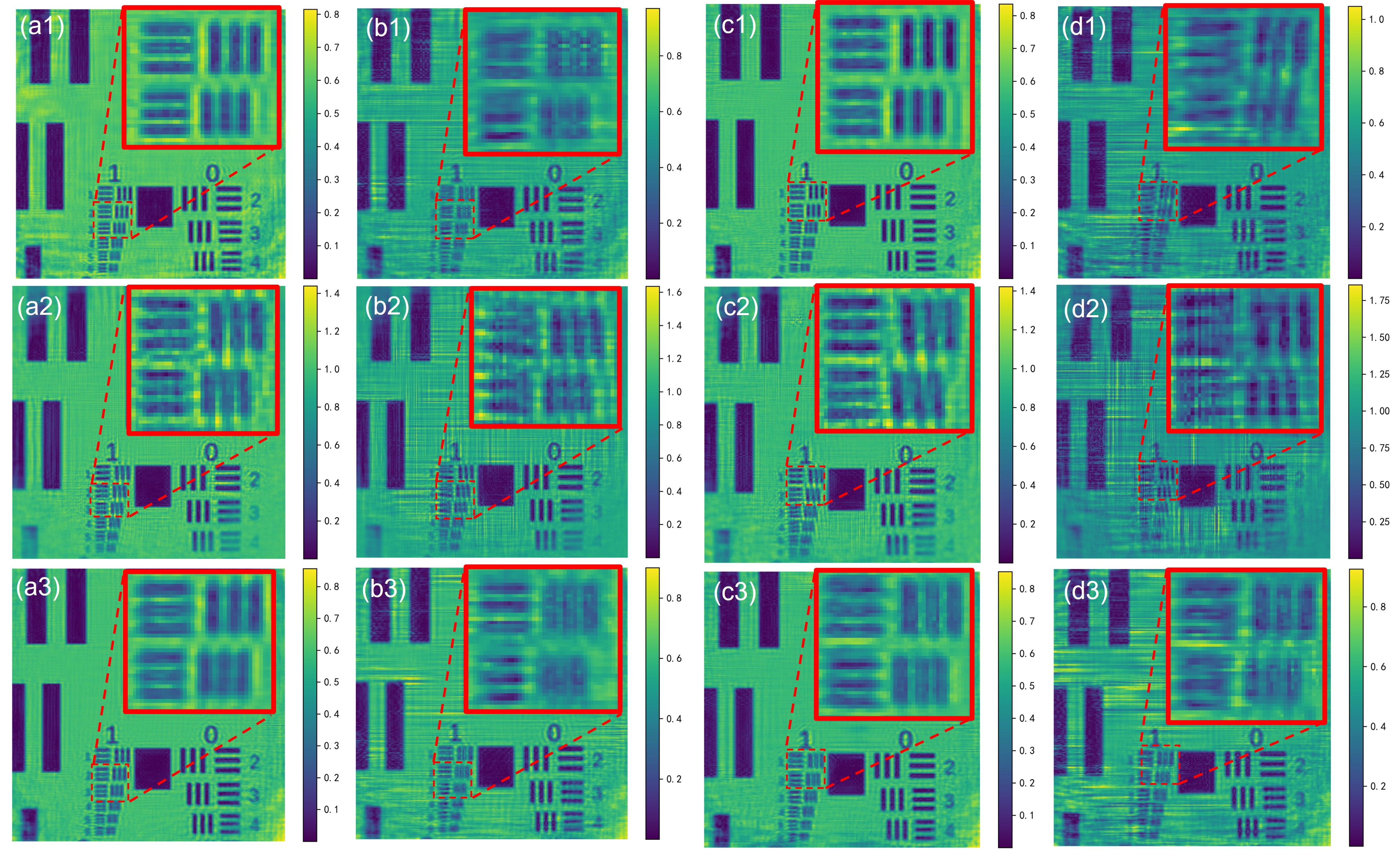}
	\caption[principle]{Comparison of reconstruction results with and without AD-based diffraction pattern shift correction under different preset pixel offsets. (a1)--(a3) and (b1)--(b3) correspond to a preset offset of $a=2$, with (a1)--(a3) showing results with shift correction and (b1)--(b3) without correction. (c1)--(c3) and (d1)--(d3) correspond to a larger offset of $a=5$, with and without shift correction, respectively. The first, second, and third rows represent the reflection model, the broadband model, and the reflection-broadband model, respectively.}
	\label{fig:fig3}
\end{figure}

The reconstruction results are illustrated in Fig.~\ref{fig:fig3}. To evaluate the robustness and stability of the proposed AD-based ptychographic algorithm under varying levels of positional misalignment, we simulate two representative scenarios with pixel offsets of $a=2$ and $a=5$, corresponding to relatively small and large positional shifts, respectively. These scenarios are shown in the first and second pairs of columns in Fig.~\ref{fig:fig3}. For each forward propagation model, a comparative analysis is performed. The first and third columns display the reconstruction results obtained by optimizing the shift parameters using automatic differentiation (AD), while the second and fourth columns present baseline reconstructions using the centrally shifted diffraction patterns without any AD-based correction. This design enables a direct comparison between corrected and uncorrected cases across both levels of misalignment. The results in Fig.~\ref{fig:fig3} clearly demonstrate that AD-based shift parameter optimization substantially improves reconstruction quality. In the corrected reconstructions, finer structural details are more accurately resolved, and prominent artifacts—especially those present in the absence of shift correction—are significantly suppressed. This improvement becomes particularly evident in the case of $a=5$, where the degradation due to uncorrected shifts is markedly more severe than in the $a=2$ scenario. These findings are consistent with our theoretical expectations, confirming the efficacy of the proposed method in mitigating the adverse effects of positional misalignments.

\begin{figure}[h!]
	\centering
	\includegraphics[width=1\linewidth]{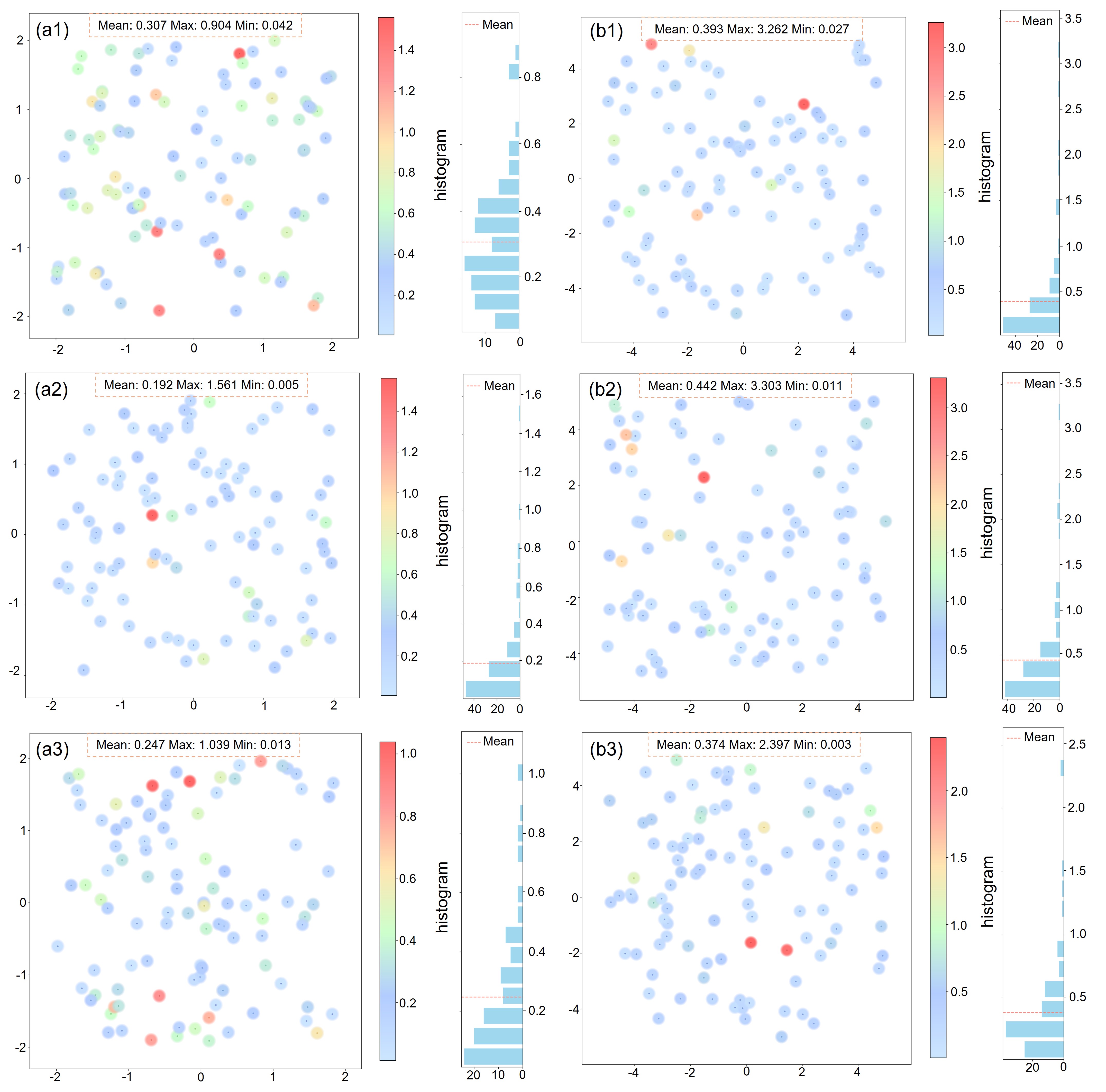}
	\caption[principle]{Evaluation of diffraction pattern shift errors using AD. (a1)-(a3) show the error distributions for preset offset of $a=2$ under the reflection model, the broadband model, and the combined reflection-broadband model, respectively. In the scatter plots, each dot represents the ground-truth shift of the diffraction pattern center, while the color indicates the deviation between the AD-calculated and true shift values, as mapped by the accompanying color bar. The corresponding histograms summarize the statistical distribution of the shift errors across all points. Similarly, (b1)-(b3) present the error results for a larger preset offset of $a=5$.}
	\label{fig:fig4}
\end{figure}

We further analyzed the residual alignment errors after correction to investigate the underlying reasons for the superior reconstruction performance. Fig.~\ref{fig:fig4} (a1)–(a3) and (b1)–(b3) present the pixel-wise deviations between the ground-truth displacements and those estimated via automatic differentiation (AD), under two levels of induced positional offsets: $a=2$ and $a=5$, respectively. In the scatter plots, each point represents a true displacement of a cropped diffraction pattern in the $x-y$ plane, with the corresponding color encoding the deviation magnitude between the estimated and actual displacement, as indicated by the color bar. Accompanying histograms summarize the statistical distribution of these deviations. For $a=2$, the majority of estimated displacements exhibit high accuracy, with errors predominantly constrained within 0.5 pixels. Even in the more challenging case of $a=5$, most estimates remain within a 1-pixel deviation, with the mean error still below 0.5 pixels. These results highlight the accuracy and robustness of the proposed AD-based displacement refinement approach, confirming its effectiveness in producing precise position estimations across a broad range of shift magnitudes.

Several additional aspects deserve attention. In Fig.~\ref{fig:fig3}, a certain degree of discrepancy is observed between the reconstructions obtained after offset correction and the ideal ground truth images. This deviation is anticipated, as positional misalignments in the diffraction patterns inherently lead to the loss of high-order diffraction components, limiting the achievable resolution. Additionally, each forward model introduces its own inherent limitations, further contributing to the reconstruction error. For instance, the reflection model exhibits anisotropic resolution, characterized by reduced fidelity along one spatial direction. In our simulation setup, the incident angle $\phi_y$ is set to non-zero value, resulting in higher resolution along the vertical axis compared to the horizontal axis. This directional imbalance is clearly evident in subfigures (a1) and (c1) of Fig.~\ref{fig:fig3}, where group number "1" shows elements "5" and "6" clearly blurred in the horizontal direction, merging into a streak, while the vertical direction retains three faint stripes, highlighting the sharper resolution along this axis. In the broadband model, reconstruction accuracy is further compromised by the numerical monochromatization process, which inherently introduces information loss and modeling error in the synthesized diffraction patterns, as previously discussed in \cite{liu2023broadband}. This degradation can be seen by comparing subfigures (a1) with (a3) and (c1) with (c3), where the reflection-broadband reconstructions consistently underperform relative to their reflection model counterparts.

\subsection{ Experiment }

\begin{figure}[h!]
	\centering
	\includegraphics[width=1\linewidth]{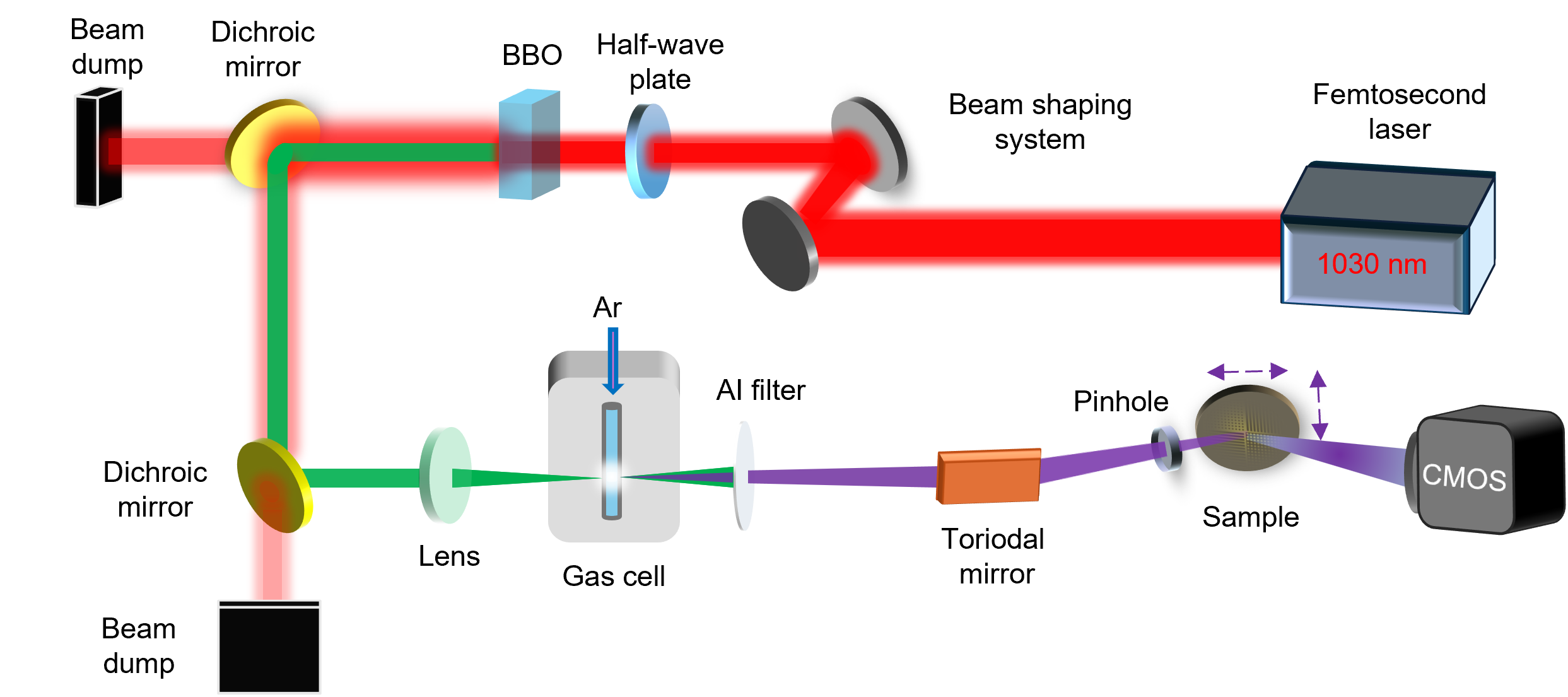}
	\caption[principle]{A schematic of the experimental setup used for ptychographic imaging. An infrared pulse from a femtosecond laser is frequency-doubled using a BBO crystal. Two dichroic mirrors (DC) are used to separate the infrared and frequency-doubled (green) light. The driving beam is then tightly focused into an argon-filled gas cell to produce high harmonic generation (HHG) extreme ultraviolet (EUV) radiation. The separation of EUV light from the driving laser is achieved through a thin aluminum metal filter. A toroidal mirror subsequently focuses the EUV beam onto the sample, in front of which a pinhole is placed to modulate the incident beam. The reflected light is finally measured by a CMOS detector. }
	\label{fig:fig5}
\end{figure}

The experimental setup, as illustrated in Fig.~\ref{fig:fig5}, employs a commercial 190 fs Yb:KGW femtosecond laser with a central wavelength of 1030 nm. The driving laser source, Pharos (Light Conversion), operates at a repetition rate of 4 kHz. A portion of its output is directed toward 515 nm light generation, with a measured pulse energy of 0.755 mJ, corresponding to an average power of 3.02 W. This portion is frequency-doubled using a 0.5 mm thick barium borate (BBO) crystal to produce 515 nm pulses with an average power of 1.66 W and a pulse duration of 180 fs. The fundamental and second harmonic beams, which co-propagate, are separated using two dichroic mirrors. The 515 nm green light is subsequently precisely focused into a gas cell filled with argon gas via a 250 mm focus lens to enhance the high harmonic generation (HHG) efficiency.

To completely block the 515 nm driving laser light, a 200 nm-thick aluminum filter was employed. The transmitted high-harmonic ultraviolet light then passes through a toroidal mirror. The toroidal mirror has a focal length of 400 mm and is aligned at an incident angle of approximately 6°. To verify the effectiveness and accuracy of the proposed method, experiments were conducted using an atomic force microscopy (AFM) testing target (HS-100MG) with better flatness and lower surface height variations. This target represents a common structure similar to those frequently observed in wafer manufacturing, where $\mathrm{SiO_2}$ layers are fabricated on $\mathrm{Si}$ substrates. To modulate the incident beam, a pinhole with a diameter of 40 \textmu m was positioned 5 mm upstream of the sample. A CMOS camera with a pixel size of 11 \textmu m was placed at a distance $z=25$ mm downstream of the sample to record the resulting diffraction patterns. Due to the cutoff region produced by the 515 nm laser and the high absorption rate of the aluminum film, the generated high-order harmonics are primarily confined to the spectral range of 39--58~nm. The dominant components within this range are the $9^\text{th}$, $11^\text{th}$, and $13^\text{th}$ harmonics, corresponding to wavelengths of 57.2 nm, 46.8 nm, and 39.6 nm, respectively. Among these, the 46.8~nm harmonic is the most intense, contributing approximately 67.4\% of the total harmonic energy, while the 57.2~nm and 39.6~nm harmonics account for around 16.8\% and 15.6\%, respectively.

In this study, we employed a Fermat spiral scanning strategy to collect diffraction patterns from a total of $K=200$ scanning positions. To facilitate subsequent experiments, each recorded diffraction pattern was initially cropped to a size of $N=512 \times 512$ pixels, roughly centered around the diffraction peak.  The incidence angle was set to approximately $\phi_y=70^{\circ}$. To ensure sufficient oversampling ratio \cite{miao2003phase} and to better constrain the probe, a pinhole was positioned in front of the sample. This physical constraint on the probe not only improved the spatial confinement but also provided a more accurate initial estimate of the probe profile compared to a random guess. Consequently, this reduced the required update step size during the probe reconstruction process.

\begin{figure}[h!]
	\centering
	\includegraphics[width=1\linewidth]{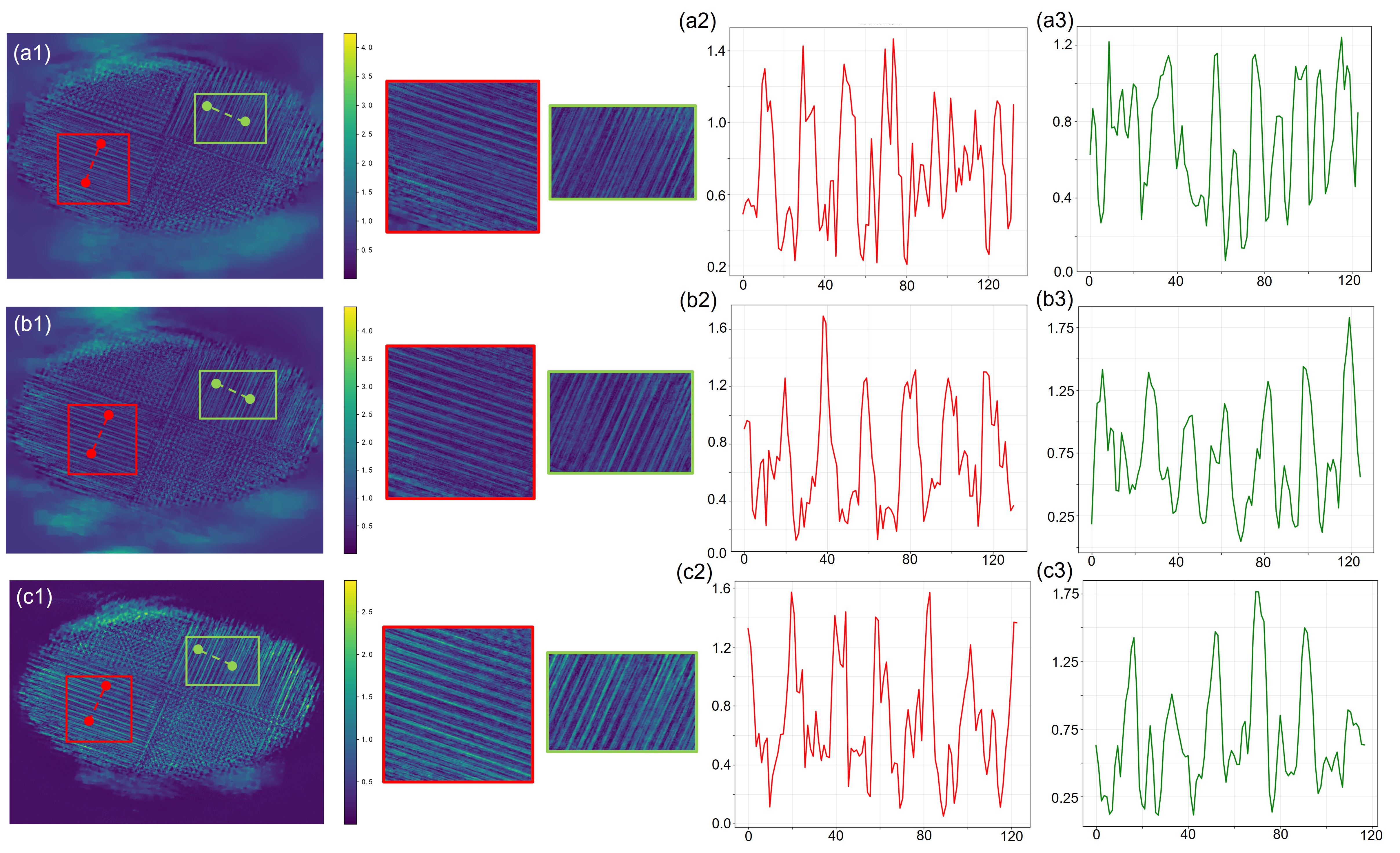}
	\caption[principle]{Comparison of ptychographic reconstruction results in EUV experiments with and without AD-based correction for diffraction pattern shifts. (a1) Reconstruction without AD-based shift correction, with magnified views of red and green boxed areas on the right. (a2) and (a3) display the intensity profiles along dashed lines in the red and green boxes, respectively. (b1) Reconstruction with AD-based shift correction, with corresponding intensity profiles in the red and green boxed regions shown in (b2) and (b3). (c1) Reconstruction from joint optimization of all variables, including diffraction pattern shifts, with intensity profiles shown in (c2) and (c3) for the red and green regions, respectively.}
	\label{fig:fig6}
\end{figure}

Given the inherent dispersion issues associated with samples in the extreme ultraviolet (EUV) regime, we did not adopt a reflection-broadband model for this experiment. Instead, we performed independent reconstructions of both the object and probe at three discrete wavelengths simultaneously, with 47.9 nm selected as the central wavelength for analysis. To evaluate the effectiveness of our AD-based diffraction shift correction algorithm, we conducted two sets of reconstructions for comparison. The first utilized the coarsely cropped diffraction patterns without any shift correction, while the second incorporated optimization of shift variables $\boldsymbol{\xi}$ with a learning rate of 0.05 to correct for misalignments after 5 epochs. The comparative results, illustrating the impact of shift correction on reconstruction quality, are presented in Fig.~\ref{fig:fig6} (a1)-(a3) and (b1)-(b3). The additional experiments presented in Fig.~\ref{fig:fig6} (c1)-(c3) are designed to demonstrate the effectiveness of our model in jointly optimizing all relevant variables.

%The experimental results in Fig.~\ref{fig:fig6} demonstrate the effectiveness of AD-based correction for diffraction pattern shifts in enhancing the quality of ptychographic reconstruction. 
Without applying AD-based positional correction, the reconstructed sample exhibits blurred features and poorly defined fringes, as shown in the central areas of the zoom-in views in Fig.~\ref{fig:fig6} (a1). In contrast, the use of AD to correct the diffraction pattern shifts results in noticeably sharper fringe structures (Fig.~\ref{fig:fig6} (b1)). This improvement is quantitatively supported by the corresponding intensity profiles: in Fig.~\ref{fig:fig6} (a2), several peaks become indistinct or merge together, while in Fig.~\ref{fig:fig6} (b2), the peaks are well-separated and clearly defined. Moreover, the contrast range improves from approximately 0.2--1.4 in Fig.~\ref{fig:fig6} (a2) to 0--1.6 in Fig.~\ref{fig:fig6} (b2). A similar trend is observed in Fig.~\ref{fig:fig6} (a3) and (b3), where the intensity range improves from 0--1.2 to 0--1.75. Considering that the grating period is 5~\textmu m and the calculated object pixel size is 207.74~nm, each grating period corresponds to approximately 24 pixels (or slightly fewer, considering the inclination of grating direction) in the reconstruction. Therefore, when comparing Fig.~\ref{fig:fig6} (a2) and (b2), as well as Fig.~\ref{fig:fig6} (a3) and (b3), the peak spacing in (b2) and (b3) aligns more closely with the actual physical period of the grating. These results confirm that AD-based shift correction significantly enhances the reconstruction clarity and structural resolution.

Further improvements are observed when all relevant variables, including diffraction pattern shifts, are jointly optimized. Although the intensity ranges in Fig.~\ref{fig:fig6} (c2) and (c3) remain comparable to those in Fig.~\ref{fig:fig6} (b2) and (b3), the peak-to-valley contrast is more pronounced. This correlates with the sharper and more distinct fringes in Fig.~\ref{fig:fig6} (c1). The enhanced clarity achieved through full-variable joint optimization highlights the advantage of integrating diffraction shift correction into the ptychographic framework, leading to superior reconstruction fidelity in complex experimental scenarios.

\section{Conculsion}
We have presented an automatic differentiation (AD)-based framework for correcting diffraction pattern shifts in ptychographic reconstruction. By incorporating shift parameters directly into the optimization process, our method enables refinement of the object, probe, and positional shifts within a unified computational framework. In numerical simulations with predefined artificial offsets, the recovered shift parameters show excellent agreement with the ground truth, with mean deviations consistently below 0.5 pixels. These results highlight the high accuracy and robustness of the AD-based correction mechanism. Furthermore, in experimental studies conducted in the extreme ultraviolet (EUV) regime, the method enhances reconstruction results by jointly optimizing other experimental variables, such as scanning positions, tilt angles, and background terms, within the same AD-based framework. This clearly demonstrates that the proposed approach is modular and extensible, allowing for the simultaneous training of all relevant variables, thereby improving the experimental reconstructions.

The proposed method is broadly compatible with various ptychographic modalities, including reflective geometries and broadband illumination, due to its differentiable and generalizable design. By eliminating manual parameter tuning, it reduces reliance on empirical adjustments and improves the reproducibility of complex imaging experiments. More broadly, this work exemplifies the potential of AD-based modeling to shift complexity from instrumentation to computation, enabling more intelligent, automated, and scalable reconstruction pipelines across the evolving landscape of computational microscopy.

%\section*{Acknowledgement}
%
%The author(s) declare that financial support was received for theresearch, authorship, and publication of this article. This work was supported by the National Key Research and DevelopmentProgram of China (2021YFB3602600);the Chinese Academy of (CAS) (GJJSTD20200009) (2018-131-S);the National Sciences Natural Science Foundation of China (NSFC) (62121003) (10010108B1339-2451) and the Beijing Municipal Science andTechnology Commission (Z221100006722008).

\begin{backmatter}

\bmsection{Funding}
National Key Research and Development Program of China (2021YFB3602600, 2024YFE0205800); Chinese Academy of Sciences (GJJSTD20200009, 2018-131-S); National Natural Science Foundation of China (62121003, 10010108B1339-2451, 62405332, 62427901); Beijing Municipal Science and Technology Commission, Adminitrative Commission of Zhongguancun Science Park (Z221100006722008).

\bmsection{Disclosures}
The authors declare no conflicts of interest.

\end{backmatter}

%%%%%%%%%%%%%%%%%%%%%%% References %%%%%%%%%%%%%%%%%%%%%%%%%

%Add references with BibTeX or manually 
%\cite{Zhang:14,OPTICA,FORSTER2007,Dean2006,testthesis,Yelin:03,Masajada:13,codeexample}.

%%%%%%%%%% If using BibTeX:
\bibliography{ref}

%%%%%%%%%% If preparing manually:
% \begin{thebibliography}{1}
% \newcommand{\enquote}[1]{``#1''}

% \bibitem{Zhang:14}
% Y.~Zhang, S.~Qiao, L.~Sun, Q.~W. Shi, W.~Huang, L.~Li, and Z.~Yang,
%   \enquote{Photoinduced active terahertz metamaterials with nanostructured
%   vanadium dioxide film deposited by sol-gel method,}
%   {\protect\JournalTitle{Optics Express}} \textbf{22}, 11070--11078 (2014).

% \bibitem{Optica}
% {Optica}, \enquote{{Optica Publishing Group},}
%   \url{http://www.opg.optica.org}.

% \bibitem{FORSTER2007}
% P.~Forster, V.~Ramaswamy, P.~Artaxo, T.~Bernsten, R.~Betts, D.~Fahey,
%   J.~Haywood, J.~Lean, D.~Lowe, G.~Myhre, J.~Nganga, R.~Prinn, G.~Raga,
%   M.~Schulz, and R.~V. Dorland, \enquote{Changes in atmospheric consituents and
%   in radiative forcing,} in \enquote{Climate Change 2007: The Physical Science
%   Basis. Contribution of Working Group 1 to the Fourth Assesment Report of
%   Intergovernmental Panel on Climate Change,}  S.~Solomon, D.~Qin, M.~Manning,
%   Z.~Chen, M.~Marquis, K.~B. Averyt, M.~Tignor, and H.~L. Miler, eds.
%   (Cambridge University Press, 2007).

% \end{thebibliography}

\end{document}